\documentclass[twocolumn,showpacs]{revtex4}

\usepackage{graphicx}%
\usepackage{dcolumn}
\usepackage{amsmath}
\input{psfig.sty}

\def\r#1{\ignorespaces $^{#1}$}

\begin{document}

\title[alphas]{Measurement of the Strong Coupling Constant \\ from Inclusive Jet Production 
at the Tevatron $\bar pp$ Collider}
\begin{abstract}
We report a measurement 
of the strong coupling constant, $\alpha_s(M_Z)$, extracted from inclusive jet 
production in $p\bar{p}$ collisions at $\sqrt{s}=$1800 GeV. The QCD 
prediction for the evolution of $\alpha_s$ with jet transverse energy $E_T$ 
is tested over the range 40$\,<\,$$E_T$$\,<\,$450 GeV using $E_T$ for the 
renormalization scale.
The data show good agreement with QCD in the region below 250 GeV.
In the text we discuss the data-theory comparison in the region from
250 to 450 GeV.
The value of $\alpha_s$ at the mass of the  $Z^0$ boson 
averaged over the range 40$\,<\,$$E_T$$\,<\,$250 GeV
is found to be $\alpha_s(M_{Z})=
0.1178 \pm 0.0001{(\rm stat)}^{+0.0081}_{-0.0095}{\rm (exp.\,syst)}$. 
The associated theoretical uncertainties are mainly due to the choice of 
renormalization scale ($^{+6\%}_{-4\%}$) and input parton  distribution functions (5\%).
\end{abstract}

\maketitle

Jet production at hadron colliders
provides an excellent opportunity for testing the theory of strong
interactions,  Quantum Chromodynamics (QCD)~\cite{QCD}. 
QCD has achieved remarkable success in describing hadron 
interactions at short distances (large momentum transfers),
owing to the property of asymptotic freedom~\cite{asf1}.
Asymptotic  freedom predicts a logarithmic decrease 
 of the  coupling strength, $\alpha_s(\mu)$, as 
the momentum scale $\mu$ characterizing a
process increases.
Processes with large momentum transfer can then be 
described by an expansion in powers of $\alpha_s(\mu)$.
The value of $\alpha_s$, a free parameter of QCD, is one of the  
fundamental constants of nature.  Its determination is the essential 
measurement of QCD, and the observation of its evolution, or {\em running}, 
with momentum transfer is one of the key tests of the theory. 
At $e^+e^-$ colliders $\alpha_s$ has  been measured from 
the fragmentation functions~\cite{jet-fragmentation-ee}, 
event shapes~\cite{event-shapes-ee-delphi},
jet production rates~\cite{jet-rates-ee-opal} and $\tau$ lepton decay
properties~\cite{tau-lepton-cleo}.
In lepton-hadron collisions, $\alpha_s$ has been measured from scaling 
violations~\cite{DIS-scaling-violations-review}, 
jet production rates~\cite{DIS-jet-rates-zeus} and 
momentum sum rules~\cite{DIS-sum-rules}.
A precise value for $\alpha_s$ 
has also been obtained from a global fit to 
 properties of the $Z^0$ boson measured at the CERN LEP and 
the SLAC SLC $e^+e^-$ colliders and the $W$ boson 
and top quark masses~\cite{Z0-global-fit}.
A review of these and other $\alpha_s$ measurements can be found 
in~\cite{Hinchliffe}. In this letter, we present a measurement of
$\alpha_s$ from the 
inclusive jet cross section in $\bar pp$ collisions over the 
jet transverse energy ($E_T$) range from 40 to 450 GeV.

This measurement is based on a data sample 
of integrated luminosity $87~{\rm pb}^{-1}$ collected
by the Collider Detector at Fermilab (CDF) during the 1994-95 
run  (Run 1b) of the Fermilab Tevatron $\bar pp$ collider operating 
at $\sqrt{s}=1.8$ TeV.
The CDF detector is described elsewhere~\cite{CDF-detector}. 
Details of the measurement of the inclusive jet differential cross 
section can be found in \cite{CDF-Run1b}. Briefly, 
jets are reconstructed using an iterative
fixed cone algorithm
 with a radius
$R={(\Delta\eta^2+\Delta\phi^2)}^{1/2}=0.7$, where 
$\eta\equiv -\ln(\mbox{tan}\frac{\theta}{2})$ is the 
pseudorapidity,   
evaluated from the angle $\theta$ between the centerline of the jet cone 
and  the proton beam line, and $\phi$ is the 
azimuthal angle. The inclusive jet cross section
includes all jets in an event  in the 
pseudorapidity range $0.1<|\eta|<0.7$.
The measured spectrum is corrected for the calorimeter response, resolution
and the underlying event energy using an iterative unsmearing 
procedure which changes both the energy scale and the normalization simultaneously.
The value of $\alpha_s$ is determined by comparing
the jet cross section  with
the next to leading order (NLO)  perturbative QCD 
predictions~\cite{Walter-alphas}.
In the $E_T$ region studied, 
the non-perturbative contributions to the inclusive jet cross section are estimated to be negligible~\cite{Ellis-rsep}. The procedure of extracting $\alpha_s$
can be summarized by the  equation
\begin{equation} 
\label{eq-main}
\frac{d\sigma}{dE_T}=\alpha_s^2(\mu_R)\hat{X}^{(0)}(\mu_F,E_T)\left [1+
\alpha_s(\mu_R)k_1(\mu_R,\mu_F,E_T)\right ]
\end{equation}
where 
$\frac{d\sigma}{dE_T}$ is the transverse energy distribution of the inclusive
jets, 
$\mu_R$ and $\mu_F$, related to $E_T$ by a scale factor, 
are the renormalization and factorization  scales, 
$\alpha_s^2(\mu_R)\hat{X}^{(0)}(\mu_F,E_T)$ is the leading order (LO) prediction 
for the inclusive jet cross section, and
$\alpha_s^3(\mu_R)\hat{X}^{(0)}(\mu_F,E_T)k_1(\mu_R,\mu_F,E_T)$
the NLO contribution.
Both $\hat{X}^{(0)}(\mu_F,E_T)$ and $k_1(\mu_R,\mu_F,E_T)$ are calculated with the 
{\sc jetrad} Monte Carlo program~\cite{JETRAD} based on the techniques 
described in \cite{Jetrad-technique-II} and the 
matrix elements of \cite{Matrix-Elements}. NLO QCD predictions
for the inclusive jet cross section are also available
in~\cite{Ellis:1990ek,Aversa:1988mm} and agree well with those of {\sc jetrad}.
All calculations are performed in the modified minimal subtraction, 
$\overline{{\rm MS}}$~, scheme\cite{MSbar-scheme}. The {\sc jetrad} Monte Carlo
program generates events with weighting factors,
so that the jet clustering algorithm and $E_T$ and  $\eta$ cuts, mimicking the 
experimental requirements, are directly applied to the final state partons.  
The jet clustering in {\sc jetrad} is governed by a cone radius $R$ and a 
phenomenological parameter 
${\cal{R}}_{sep}$(default value =1.3), introduced to match the 
experimental efficiency of identifying overlapping jets\cite{Ellis-rsep}. 
If two partons are more 
than ${\cal{R}}_{sep}\times R$ apart, they are identified as two
distinct jets, otherwise they are merged into a single jet. 

\begin{figure}
\centerline{\psfig{figure=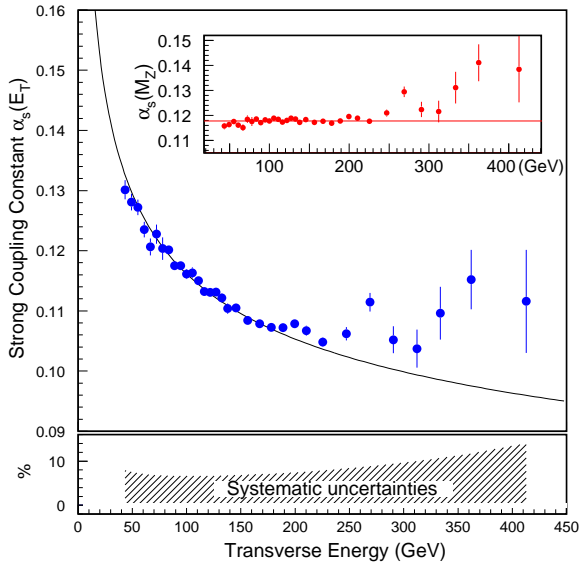,width=3.3in}}
\caption{The strong coupling constant as a function of $E_T$  
for $\mu_R=E_T$ measured using {\sc cteq4m} parton distributions. 
The shaded area shows  the experimental systematic uncertainties. 
The curved line represents the NLO QCD prediction for the evolution of 
$\alpha_s(E_T)$ using $\alpha_s(M_Z)=0.1178$, the average value 
obtained in the region $40<E_T<250$ GeV.
The $\alpha_s(M_Z)$ extracted from $\alpha_s(E_T)$ is shown in the inset
along with the weighted average as the horizontal line.}
\label{alphas_main}
\end{figure}

The inclusive jet data are divided into 33 $E_T$ bins, 
from which we obtain  statistically independent measurements of
$\alpha_s$ for 33 different values of $\mu_R$. 
The $\alpha_s$ values derived for $\mu_R=\mu_F=E_T$ using 
{\sc cteq4m}~\cite{CTEQ4M} parton distribution functions (PDFs) 
are presented in Fig.~\ref{alphas_main}. For $E_T$$<$250 GeV, there is 
good agreement with QCD
predictions for the  running of the coupling constant.
The behavior of $\alpha_s$ at higher  $E_T$ values  
is a direct reflection of the 
excess observed  in the inclusive jet cross section~\cite{CDF-Run1b}. 
The discrepancy with the NLO QCD predictions
in this region, though not well understood, may be accommodated by the 
flexibility allowed by the world data in determining the high-$x$ 
gluon component in the parton distributions~\cite{CTEQ4M}.

The measured values of $\alpha_s(\mu_R)$ are evolved to the  
mass of the $Z^0$ boson, $M_Z$,
by using the solution to the 2-loop renormalization group equation 
\begin{eqnarray}
\alpha_s(M_Z)=\frac{\alpha_s(\mu_R)}
{1-\alpha_s(\mu_R)(b_0+b_1\alpha_s(\mu_R))\ln(\mu_R/M_Z)}
\label{eq:rge}
\end{eqnarray}
\begin{eqnarray}
b_{0}=\frac{33-2n_f}{6\pi}\hspace*{1.0cm}
b_{1}=\frac{306-38n_f}{24{\pi}^2},
\end{eqnarray}
where $n_f$ is the number of active flavors, which is equal to five (six) for 
$\mu_R$ smaller (larger) than the top quark mass. The values of 
$\alpha_s(M_Z)$ for all 33 measurements are shown in the inset of 
Fig.~\ref{alphas_main}.
Averaging over the range 40-250 GeV, we obtain
\[\alpha_s(M_Z)=0.1178\pm0.0001{\rm (stat)}.\]
Inclusion of the data with $E_T>$250 GeV results in an increase of 
the average value by  0.0001.

The running of $\alpha_s$ is tested by 
verifying if
$\alpha_s(M_Z)$ is independent of the energy scale $E_T$ at which 
the jet cross section is measured. 
The 27 values of $\alpha_s(M_Z)$ obtained from the data in the jet $E_T$ range 
40-250 GeV are fit to the linear function $P_0+P_1\times (E_T/E^0_T-1)$
with $E_T^0=92.8$ GeV.
The fit yields $P_0=0.1176\pm0.0003$ and $P_1=0.0003\pm0.0003$ with $\chi^2/d.o.f.=1.3$,
showing that $\alpha_s(M_Z)$ is independent of $E_T$ within
one standard deviation. This non-trivial result 
demonstrates the correctness of the QCD prediction for the 
evolution of $\alpha_s$ over the above range.

\begin{figure}
\centerline{\psfig{figure=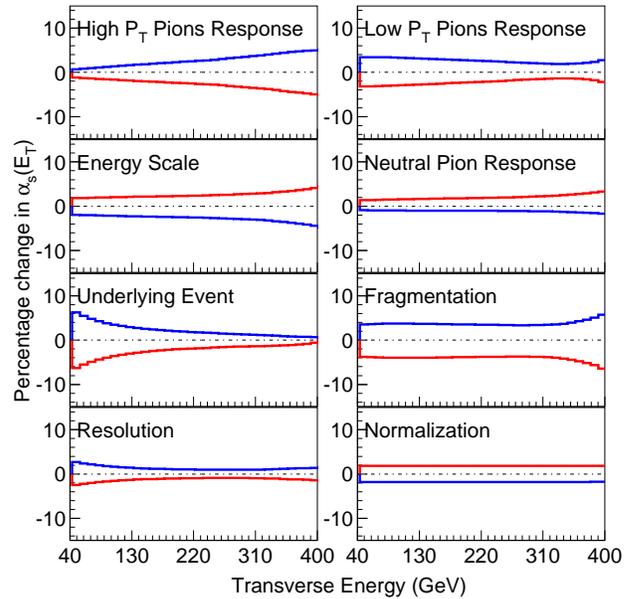,width=3.5in}}
\caption{Experimental systematic uncertainties for $\alpha_s$ measurement  
(the lines are 1 standard deviation contours), 
with
{\sc cteq4m} as input PDF and $\mu_R=\mu_F=E_T$.}
\label{exp-sys}
\end{figure}

The  experimental systematic uncertainties on the
value of $\alpha_s(M_Z)$ are derived from  
those on the inclusive jet cross section. For each source 
of systematic uncertainty described below, 
except normalization,
the inclusive jet cross
section was re-evaluated by varying the corresponding parameter
in the detector response by 1 $\sigma$.  For the normalization uncertainty
it was changed by a scale factor~\cite{CDF-Run1b}. These uncertainties
were propagated to $\alpha_s(M_Z)$ by repeating the 
procedure described above using the spectra given in
Table VI of Ref.~\cite{CDF-Run1b}. The results are shown in  
Fig.~\ref{exp-sys}.
The  deviations of $\alpha_s(M_Z)$ for each spectrum from the 
central value  are given in Table~\ref{table1}.
\begin{table}
\caption{Experimental systematic uncertainties on     
$\alpha_s(M_{Z})$ extracted using {\sc cteq4m} parton distribution 
functions.}
\label{table1}
\begin{tabular}{lcc}
\hline
\hline
\multicolumn{1}{l}
{Source of Uncertainty}            &
\multicolumn{1}{c}
{$\Delta\alpha_s$}  &
\multicolumn{1}{r}
{($\frac{\Delta\alpha_s}{\alpha_s})$ \%}   \\
\vspace{-0.2cm} \\
\hline\hline                
\vspace{-0.3cm} \\
Calorimeter high $P_T$ pion response     
&$^{+0.0036}_{-0.0055}$    & $^{+3.1}_{-4.7}$ \\              
Calorimeter low $P_T$ pion response      
&$^{+0.0027}_{-0.0033}$    & $^{+2.3}_{-2.8}$ \\              
Energy scale                             
&$^{+0.0030}_{-0.0030}$    & $^{+2.6}_{-2.6}$ \\ 
Neutral pion response        
&$^{+0.0016}_{-0.0021}$    & $^{+1.4}_{-1.8}$ \\              
Underlying event energy    
&$^{+0.0025}_{-0.0027}$    & $^{+2.1}_{-2.3}$ \\              
Jet fragmentation functions              
&$^{+0.0046}_{-0.0044}$    & $^{+3.9}_{-3.7}$ \\              
Jet energy resolution                    
&$^{+0.0015}_{-0.0017}$    & $^{+1.3}_{-1.4}$ \\              
Normalization                            
&$^{+0.0022}_{-0.0023}$    & $^{+2.0}_{-1.9}$ \\ 
\hline
\hline
\end{tabular}
\end{table}

The dominant experimental 
systematic uncertainty in the inclusive jet cross section
measurement is due to the calorimeter response to jets.
The detector response and jet energy corrections are derived from a 
combination of test-beam data and Monte-Carlo simulations.
The calorimeter response to charged pions 
was evaluated separately for  high and low transverse momentum ($P_T$) pions 
from test beam data and isolated charged tracks from $p\bar{p}$ data
with an uncertainty of $\pm 5\%$ for $P_T\le 5$ GeV,
$\pm 3\%$ for  5 GeV$<$$P_T$$<$15 GeV and $^{+3.6}_{-2.0}\%$ for $P_T\ge15$ GeV. 
During the run, the calorimeter response was monitored by using muons,
isolated particles and the measured inclusive jet cross section.
The response was found to be stable within $\pm 1\%$.
The electromagnetic calorimeter was calibrated using electrons
from $p\bar p$ interactions. The associated uncertainty, 
labeled in Fig.~\ref{exp-sys} as neutral pion response,
arises from the modeling of calorimeter response to very low energy electrons. 
The underlying event energy (non-jet energy contribution to the 
jet $E_T$)  was measured 
from  minimum bias data and its mean value was varied by
$\pm 30\%$ to evaluate the effect on the jet cross section.
The error from the jet fragmentation functions is due to the extrapolation 
of the track momentum and multiplicity distribution to the high $E_T$ region 
and from uncertainties in the track reconstruction efficiency.
The detector jet energy response has a Gaussian shape with exponential tails
on both the high and low sides and a resolution with an uncertainty of 
$\pm 10\%$.
Finally, the overall  normalization of the 
inclusive jet cross section has an uncertainty of $\pm 4.5\%$, 
dominated by the contribution from the total cross section measurement.
Summing in quadrature all the above uncertainties after 
propagation to  $\alpha_s(M_Z)$  yields  
 a total experimental systematic uncertainty of $\pm^{0.0081}_{0.0095}$.

\begin{figure}
\centerline{\psfig{figure=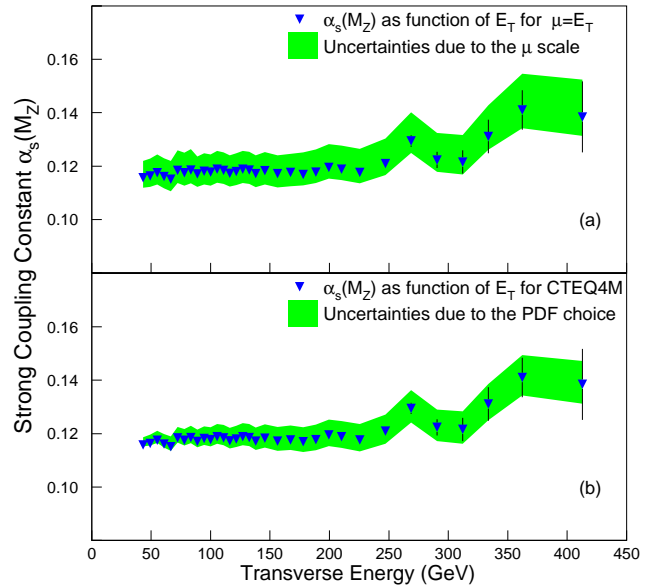,width=3.5in}}
\caption{Uncertainties in $\alpha_s(M_Z)$ due 
to the renormalization scale $\mu$, 
(a), and parton distribution
functions, (b). 
}
\label{mz_theor_plot}
\end{figure}

The theoretical uncertainties are mainly due to the
choice of renormalization and factorization scales
and parton distribution functions. The scales $\mu_F$ and $\mu_R$ 
are expected to be of the same order as the characteristic scale of
the process, which in this case is the jet $E_T$. 
We have evaluated the changes in $\alpha_s(M_Z)$ resulting from 
independently varying  $\mu_F$ and  $\mu_R$ from 
$E_T/2$ to $2E_T$ and found that the largest changes occur for $\mu_R=\mu_F$.  
For all results presented in this letter the two scales were set equal.
The sensitivity of the measured $\alpha_s(M_Z)$ to
changes in these scales is indicated by the shaded band in
Fig.~\ref{mz_theor_plot}(a).
Over the $E_T$ range from 40 to 250 GeV, the 
shift in $\alpha_s(M_Z)$ induced by changing the scale 
from $E_T/2$ ($\alpha_s(M_Z)=0.1129\pm 0.0001$)~\cite{footnote} to $2E_T$ ($\alpha_s(M_Z)=0.1249\pm 0.0001$) is
approximately $^{+6\%}_{-4\%}$, independently of $E_T$.

The coefficients $\hat{X}_0$ and $k_1$ in Eq.(~\ref{eq-main}) depend on the
PDFs,  which are obtained from global fits 
to deep inelastic scattering (DIS), Drell-Yan production 
and other collider data, including
the  inclusive jet results from Tevatron. Each PDF set has 
an associated strong coupling constant, $\alpha_s^{PDF}$.
The gluon PDF $(G(x))$ determined from the fit is highly 
    correlated with $\alpha_s^{PDF}$  because in equations describing 
    all the processes used, the $G(x)$ is always  
    accompanied by $\alpha_s^{PDF}$.
To calculate the above  coefficients, the PDFs 
are evolved using $\alpha_s^{PDF}$. 
For this procedure of measuring $\alpha_s$ to be valid,
the extracted value of $\alpha_s$ should be consistent with the 
input $\alpha_s^{PDF}$, although not necessarily equal. 
The variation in parton distributions, especially in the gluon 
distribution, allowed by the world data  
was studied by the CTEQ Collaboration by fixing the 
value of $\alpha_s^{PDF}$ to 
0.110, 0.113, 0.116, 0.119 or 0.122, with resulting
$\chi^2$ of 1388, 1323, 1323, 1388 or 1543
for  1297 non-jet data points~\cite{CTEQ4M}.
We use the {\sc cteq4a} series to study the
$\alpha_s(M_Z)$ dependence on the PDFs. 
In addition, we have studied $\alpha_s(M_Z)$ using 
PDF sets which do not include Tevatron jet results, such as the
{\sc mrst}(g$\uparrow$) set~\cite{MRST}, 
 the {\sc mrsa}$^\prime$ series~\cite{MRSA}, 
and two {\sc mrs-r} sets~\cite{MRSR}. 
The $\chi^2$, calculated by
comparing the data with the theoretical prediction in
the restricted range of 40-250 GeV, is used to quantify 
the agreement.
The minimal  $\chi^2/d.o.f.=1.38$ is obtained for {\sc cteq4m} 
($\alpha_s^{PDF}$=0.116), 
therefore we use this PDF in our final fit.
Excluding the PDFs which have obvious  disagreement
($\chi^2/d.o.f.\ge 5$), we estimate the uncertainty on the
$\alpha_s(M_Z)$ due to the PDF choice to be $\pm 5\%$. 

Finally, the variation of ${\cal{R}}_{sep}$, the jet clustering parameter, 
 from 1.3 to 2.0
results in a 5-7\% normalization change of the
inclusive jet cross section.
The corresponding variation
in the $\alpha_s(M_Z)$ measurement is 2-3\%.

To explore the flexibility in the gluon distribution at high $E_T$,
a special PDF set, {\sc cteq4hj}, was generated by
including CDF jet data  in the global fit with higher statistical weight
assigned to high $E_T$ points and a new parameterization of
the gluon distribution~\cite{CTEQ4M}. This PDF yields good 
agreement between Tevatron data and theoretical predictions.  Using this set, we
obtain $\alpha_s(M_Z)=0.1185\pm 0.0001$, averaged over the entire $E_T$ range. 

In conclusion, we have tested the evolution of the strong coupling constant 
$\alpha_s$ using the inclusive jet cross section data from 
$\bar pp$ collisions at $\sqrt{s}=1800$ GeV.
Our results 
demonstrate that for $E_T$ in the range of 40-250 GeV with $\mu_R=E_T$
the running of $\alpha_s$ is in  good agreement with 
QCD predictions.
The value of $\alpha_s$ expressed at the $Z^0$ boson mass is found to be
\[\alpha_s(M_{Z})=0.1178\pm0.0001({\rm stat})^{+0.0081}_{-0.0095}(
{\rm exp.syst})\]
This value is in good agreement with the world average 
 $\alpha_s(M_Z)=0.1181\pm 0.0020$ \cite{world}.
The theoretical uncertainties associated with 
the choice of parton distribution functions ($\sim 5\%$) and the choice of the 
renormalization scale ($^{+6\%}_{-4\%}$) are comparable to 
the experimental systematic uncertainty. 

We would like to thank W. Giele for providing the {\sc jetrad} 
program  and for 
helpful discussions and comments on all stages of this work.
We thank the Fermilab staff and the technical staffs of the
participating institutions for their vital contributions.  This work was
supported by the U.S. Department of Energy and National Science Foundation;
the Italian Istituto Nazionale di Fisica Nucleare; the Ministry of Education,
Science, Sports and Culture of Japan; the Natural Sciences and Engineering 
Research Council of Canada; the National Science Council of the Republic of 
China; the Swiss National Science Foundation; the A. P. Sloan Foundation; the
Bundesministerium fuer Bildung und Forschung, Germany; the Korea Science 
and Engineering Foundation (KoSEF); the Korea Research Foundation; and the 
Comision Interministerial de Ciencia y Tecnologia, Spain.
\bibliography{alphas_hep}%

\begin{thebibliography}{34}
\bibitem[{\citenamefont{The $\alpha_s(M_Z)$ value quoted in 
             preprint by \cite{Mesropian} 
     was obtained with $\mu_R=E_T/2$, however for this letter 
   we repeated the analysis with $\mu_R=E_T$.}}]{footnote}
\bibinfo{author}{\bibnamefont{The $\alpha_S(M_Z)$ value  quoted in preprint 
by {\bibfnamefont{C.} \bibnamefont{Mesropian}},
  \bibinfo{journal}{FERMILAB-CONF-99-177-E}  (\bibinfo{year}{1999}),
     was obtained with $\mu_R=E_T/2$, however for this letter 
   we repeated the analysis with $\mu_R=E_T$}.}
\end{thebibliography}
\newpage
\begin{widetext}
\font\eightit=cmti8
\def\r#1{\ignorespaces $^{#1}$}
\hfilneg
\begin{sloppypar}
\noindent
T.~Affolder,\r {23} H.~Akimoto,\r {45}
A.~Akopian,\r {37} M.~G.~Albrow,\r {11} P.~Amaral,\r 8  
D.~Amidei,\r {25} K.~Anikeev,\r {24} J.~Antos,\r 1 
G.~Apollinari,\r {11} T.~Arisawa,\r {45}
% ...
\end{sloppypar}
\vskip .026in
\begin{center}
(CDF Collaboration)
\end{center}

\vskip .026in
\begin{center}
\r 1  {\eightit Institute of Physics, Academia Sinica, Taipei, Taiwan 11529, 
Republic of China} \\
\r 2  {\eightit Argonne National Laboratory, Argonne, Illinois 60439} \\
\r 3  {\eightit Istituto Nazionale di Fisica Nucleare, University of Bologna,
I-40127 Bologna, Italy} \\
\r 4  {\eightit Brandeis University, Waltham, Massachusetts 02254} \\
\r 5  {\eightit University of California at Davis, Davis, California  95616} \\
\r 6  {\eightit University of California at Los Angeles, Los 
Angeles, California  90024} \\  
\r 7  {\eightit Instituto de Fisica de Cantabria, CSIC-University of Cantabria, 
39005 Santander, Spain} \\
\r 8  {\eightit Enrico Fermi Institute, University of Chicago, Chicago, 
Illinois 60637} \\
\r 9  {\eightit Joint Institute for Nuclear Research, RU-141980 Dubna, Russia}
\\
\r {10} {\eightit Duke University, Durham, North Carolina  27708} \\
\r {11} {\eightit Fermi National Accelerator Laboratory, Batavia, Illinois 
60510} \\
\r {12} {\eightit University of Florida, Gainesville, Florida  32611} \\
\r {13} {\eightit Laboratori Nazionali di Frascati, Istituto Nazionale di Fisica
               Nucleare, I-00044 Frascati, Italy} \\
\r {14} {\eightit University of Geneva, CH-1211 Geneva 4, Switzerland} \\
\r {15} {\eightit Glasgow University, Glasgow G12 8QQ, United Kingdom}\\
\r {16} {\eightit Harvard University, Cambridge, Massachusetts 02138} \\
\r {17} {\eightit Hiroshima University, Higashi-Hiroshima 724, Japan} \\
\r {18} {\eightit University of Illinois, Urbana, Illinois 61801} \\
\r {19} {\eightit The Johns Hopkins University, Baltimore, Maryland 21218} \\
\r {20} {\eightit Institut f\"{u}r Experimentelle Kernphysik, 
Universit\"{a}t Karlsruhe, 76128 Karlsruhe, Germany} \\
\r {21} {\eightit Center for High Energy Physics: Kyungpook National
University, Taegu 702-701; Seoul National University, Seoul 151-742; and
SungKyunKwan University, Suwon 440-746; Korea} \\
\r {22} {\eightit High Energy Accelerator Research Organization (KEK), Tsukuba, 
Ibaraki 305, Japan} \\
\r {23} {\eightit Ernest Orlando Lawrence Berkeley National Laboratory, 
Berkeley, California 94720} \\
\r {24} {\eightit Massachusetts Institute of Technology, Cambridge,
Massachusetts  02139} \\   
\r {25} {\eightit University of Michigan, Ann Arbor, Michigan 48109} \\
\r {26} {\eightit Michigan State University, East Lansing, Michigan  48824} \\
\r {27} {\eightit University of New Mexico, Albuquerque, New Mexico 87131} \\
\r {28} {\eightit The Ohio State University, Columbus, Ohio  43210} \\
\r {29} {\eightit Osaka City University, Osaka 588, Japan} \\
\r {30} {\eightit University of Oxford, Oxford OX1 3RH, United Kingdom} \\
\r {31} {\eightit Universita di Padova, Istituto Nazionale di Fisica 
          Nucleare, Sezione di Padova, I-35131 Padova, Italy} \\
\r {32} {\eightit University of Pennsylvania, Philadelphia, 
        Pennsylvania 19104} \\   
\r {33} {\eightit Istituto Nazionale di Fisica Nucleare, University and Scuola
               Normale Superiore of Pisa, I-56100 Pisa, Italy} \\
\r {34} {\eightit University of Pittsburgh, Pittsburgh, Pennsylvania 15260} \\
\r {35} {\eightit Purdue University, West Lafayette, Indiana 47907} \\
\r {36} {\eightit University of Rochester, Rochester, New York 14627} \\
\r {37} {\eightit Rockefeller University, New York, New York 10021} \\
\r {38} {\eightit Rutgers University, Piscataway, New Jersey 08855} \\
\r {39} {\eightit Texas A\&M University, College Station, Texas 77843} \\
\r {40} {\eightit Texas Tech University, Lubbock, Texas 79409} \\
\r {41} {\eightit Institute of Particle Physics, University of Toronto, Toronto
M5S 1A7, Canada} \\
\r {42} {\eightit Istituto Nazionale di Fisica Nucleare, University of Trieste/
Udine, Italy} \\
\r {43} {\eightit University of Tsukuba, Tsukuba, Ibaraki 305, Japan} \\
\r {44} {\eightit Tufts University, Medford, Massachusetts 02155} \\
\r {45} {\eightit Waseda University, Tokyo 169, Japan} \\
\r {46} {\eightit University of Wisconsin, Madison, Wisconsin 53706} \\
\r {47} {\eightit Yale University, New Haven, Connecticut 06520} \\
\r {(\ast)} {\eightit Now at Carnegie Mellon University, Pittsburgh,
Pennsylvania  15213} \\
\r {(\ast\ast)} {\eightit Now at Northwestern University, Evanston, Illinois 
60208} \\
\r {(\ast\ast\ast)} {\eightit Now at University of California, Santa Barbara, CA
93106}
\end{center}

\end{widetext}
\end{document}